\documentclass{aa}
\usepackage{graphicx}
\begin{document}

\title{\bf RR Lyrae variables in Galactic globular clusters}
\subtitle{\bf IV. Synthetic HB and RR Lyrae predictions. }

\author{S. Cassisi \inst{1}, M. Castellani \inst{2}  \and  F. Caputo \inst{2}  \and  V. Castellani  \inst{2}$^,$\inst{3}}
\offprints {M. Castellani, \email {m.castellani@mporzio.astro.it}}
\institute{INAF, Osservatorio Astronomico di Teramo,
via Mentore Maggini, 64100 Teramo, Italy
 \and INAF,
Osservatorio Astronomico di Roma, via Frascati 33, 00040 Monte
Porzio Catone, Italy \and INFN Sezione di Ferrara,via Paradiso 12,
44100 Ferrara,Italy}

\date{Received ; accepted }





%



\abstract{We present theoretical predictions concerning horizontal
branch stars in globular clusters, including RR Lyrae variables,
as derived from synthetic procedures collating evolutionary and
pulsational constraints. On this basis, we explore the predicted
behavior of the pulsators as a function of the horizontal branch
morphology and over the metallicity range $Z$=0.0001 to 0.006,
revealing an encouraging concordance with the observed
distribution of fundamentalised periods with metallicity.
Theoretical relations connecting periods to $K$ magnitudes and
$BV$ or $VI$ Wesenheit functions are presented, both appearing
quite independent of the horizontal branch morphology only with
$Z\ge$ 0.001. Predictions concerning the parameter R are also
discussed and compared under various assumptions about the
horizontal branch reference luminosity level.

}

   \maketitle


\section{Introduction}

The pulsational variability of RR Lyrae stars in Galactic globular
clusters has represented for several decades  intriguing
evidence which has stimulated a large number of investigations. It
was, indeed, early understood that the pulsation is governed by the
physical structure of the stellar objects, providing 
independent access to the evolutionary features of low mass metal
poor stellar structures.

In the first paper of this series (Castellani, Caputo \&
Castellani 2003), we have revisited the present status of the art,
presenting and discussing on purely empirical grounds the current
observational scenario. Paper II (Marconi et al. 2003)  tests the pulsational predictions for low mass
structures with $Z$=0.001, calibrating the theory on the rich
sample of RR Lyrae stars in the Galactic globular M3, while
Paper III (Di Criscienzo, Marconi \& Caputo 2004) dealt with
pulsational predictions covering the whole range of mass and metal
content, as  expected in the Galactic globular cluster family.

On this basis, we are now in the position to attempt a direct
connection between pulsation and evolution theories by
investigating the predicted behavior of suitable low mass stellar
models evolving through the central He burning, horizontal branch (HB) evolutionary phase. We will perform this task making use of the
well-known  synthetic HB (SHB) procedure, i.e. by using 
evolutionary theory to  predict the distribution of HB stars in
globular clusters of selected metallicities. Adopting the
instability strip boundaries given in Paper III, from each
computed SHB one easily derives the number and the period of
predicted RR Lyrae pulsators, to be compared with the
observational scenario presented in Paper I.

As in Paper I, in this paper we will remove the problem of the
pulsation mode by considering only fundamentalised periods ($P_f$)
to be compared with  similar observational data. The problem of
pulsational modes  will be addressed in forthcoming papers, where
we will approach also the discussion of RR Lyrae stars in selected
clusters, such as the peculiar case of NGC6441.

\section{Synthetic Horizontal Branches}

The main ingredient to produce SHBs is obviously given by suitable
sets of HB stellar models. To this purpose, we computed a fine
grid of HB models for the selected metallicities $Z$=0.0001,
0.0003, 0.0006, 0.001, 0.003 and 0.006 assuming an original He
content $Y=0.23$, implemented with a grid  for $Z$=0.006 but
$Y$=0.245  to investigate the effects of the Galactic correlation
between He and metals. All computations start from Zero Age HB
(ZAHB) structures as obtained, for each given metallicity and
helium content, by adopting the helium core mass and the envelope
chemical abundance profile suitable for a Red Giant Branch (RGB)
progenitor with mass equal to $0.8 M_\odot$, with 5\% by mass of
$^{12}$C added in the core to account for the nucleosynthesis
during the He flash. The models have been followed through the
whole phase of central He burning and He shell burning and, for
those reaching the Asymptotic Giant Branch (AGB) phase, until the
onset of thermal pulses.

The models make use of OPAL radiative opacities (Iglesias~\&
Rogers 1996) for temperatures higher than $10000$ K, while for
lower temperatures,  molecular opacity tables provided by
Alexander~\& Ferguson (1994) have been adopted. Both high and
low-temperature opacities have been computed by assuming a solar
scaled heavy element distribution (Grevesse 1991). The equation of
state (EOS) is from Straniero (1988), supplemented at lower
temperatures by a Saha EOS. The outer boundary conditions have
been fixed according to the $T(\tau)$ relation given by
Krishna-Swamy (1966). Concerning the treatment of the
superadiabatic layers, the mixing-length calibration provided by
Salaris~\& Cassisi (1996) has been adopted. Other physical inputs
are the same as in Cassisi~\& Salaris (1997).

\begin{figure}
\centering
\includegraphics[width=8cm]{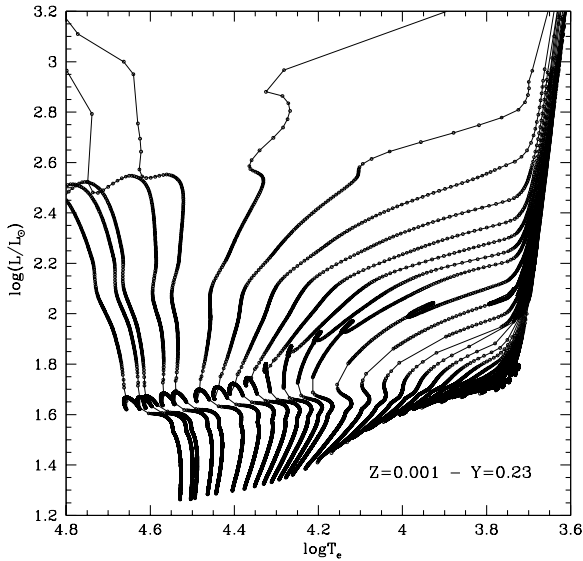}
\caption {The set of HB evolutionary tracks for metallicity
$Z$=0.001. Dots along the tracks mark the points used for  the
interpolation allowing to predict  the luminosity and  effective
temperature of a HB model for each given value of mass and HB
evolutionary time.}
\end{figure}

\begin{figure}
\centering
\includegraphics[width=8cm]{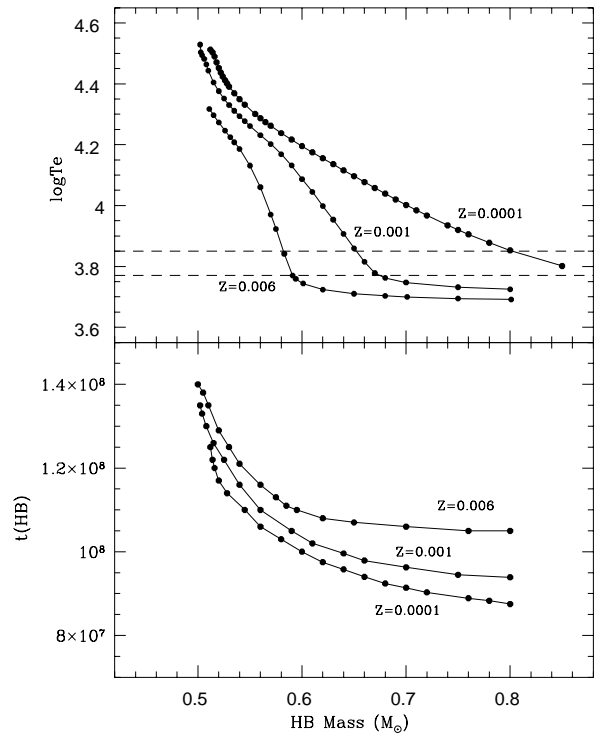}
\caption {Upper panel: The effective temperature of ZAHB models as
a function of the stellar mass for three values of the metallicity
$Z$. Dashed lines give an indication of the range of temperatures
covered by the instability strip. Lower panel: as in the upper
panel, but for the evolutionary times in the central He burning
phase.}
\end{figure}

As an example, Fig. 1 shows the data set of evolutionary tracks
for $Z$=0.001, where dots along the tracks mark the points used in
the the interpolation which allows us to predict luminosity and
effective temperature of a HB model for each given value of the
mass and HB evolutionary time. Tables containing the whole
set of HB evolutionary lines are available at
http://www.mporzio.astro.it/$\sim$marco/sintetici/, together with
detailed results of our synthetic HB simulations.
 Figure 2 shows the mass - effective temperature relation for ZAHB
models at three selected metallicities. The figure reveals the
well known occurrence in which for increasing the metallicity the
ZAHB models tend to accumulate at the lower effective temperatures
so that it decreases the range of masses predicted within the RR
Lyrae instability strip (approximately, 3.77$\le$ log$T_e \le$
3.85). The lower panel in the same figure gives the central He
burning lifetimes again as a function of  ZAHB masses and for the
three selected metallicities.

Making use of the evolutionary models, synthetic HBs have been
produced by assuming a Gaussian random distribution of masses
centered on a central mass $\langle M_{HB} \rangle$ together with
a flat random distribution of post-ZAHB evolutionary time.
According to the adopted procedure, a synthetic HB provides for
each star its mass, luminosity and effective temperature. Adopting
atmosphere models by Castelli, Gratton \& Kurucz (1997), we
finally evaluate $U,B,V,R,I,J$ and $K$ magnitudes for all the
objects. Moreover, based on the edges of the pulsation region
given in Paper III, we determine whether a star is crossing the RR
Lyrae instability strip. As a result, for each given total number
of stars in the He burning evolutionary phase, we eventually
obtain the number of HB stars to the blue (B), within (V) or to
the red (R) of the instability strip, together with the predicted
number of stars in the early AGB phase. In this way, we evaluate
the so-called "HB type" (Lee 1990), as given by the ratio
(B-R)/(B+V+R), and the number ratio R$^{AGB}$ between AGB and HB
stars. For the models falling within the pulsation region, i.e.,
for RR Lyrae candidates, we derive the predicted mean visual
magnitude $\langle M_V^{RR}\rangle$, together with their period
frequency histogram and the value of the mean fundamental period
$\langle$log$P_f\rangle$, as evaluated according to the relation
given in Paper III.

Comparison with parallel evolutionary tracks for RGB structures,
as computed for the various assumptions on $Z$, allows finally to
predict the value of the parameter R (Iben 1968), as given by the
number ratio of HB to RGB stars  more luminous than the HB
luminosity level. According to the different choices made in the
literature, this last quantity is alternatively defined as the
bolometric magnitude of the ZAHB at the mean temperature of the
instability strip log$T_e$=3.83 (R$^I$), as the absolute visual
magnitude of the faintest RR  Lyrae pulsators (R$^{II}$) or the
mean RR Lyrae $\langle M_V^{RR}\rangle$ magnitude (R$^{III}$).

However, before proceeding one has to take into consideration the
already known small discrepancy between evolutionary and
pulsational predictions. As discussed in  Caputo et al. (1999, but
see also Paper II)  a straightforward application of pulsational
results to HB evolutionary models predicts periods slightly larger
than observed in well studied clusters like M5 and M3. Using M3 as
a test, we now find that with $Z\sim$ 10$^{-3}$ the predictions of
our SHB computation would give the right interval $\Delta$log$P_f$
= log $P_f^{max}$ - log P$_f^{min}$ (see Paper I), but with the
period frequency histogram slightly shifted toward larger values.
On this basis only, one would conclude that the adopted
pulsational theory gives the right width in temperature of the
instability strip, but either the predicted strip is too cool, or
the models are too luminous, if not a combination of both.

In Paper II, we have presented what we regard as the reasonable
indication that the dilemma should be solved in the sense that
current evolutionary HB models are indeed too luminous, by an
amount which depends on the input physics adopted in the various
available evolutionary computations. Here we add that numerical
experiments have shown that adopting the most recent OPAL EOS our
HB models will become even more luminous. We feel that this
indicates that the overluminosity is not a matter of EOS, but very
likely of uncertainties in the sophisticated mechanisms governing
the mass size of the  He core in the Red Giant progenitors, such
cooling by neutrinos and electron conduction. So, we decided to
retain the already quoted theoretical framework, but calibrating
the luminosity of the models to reproduce the period distribution
as observed in M3. On this basis, the luminosity of our He burning
models has been decreased by $\Delta$log$L$=0.04. Such an
empirical correction leaves substantially unchanged the
evolutionary scenario, allowing a close agreement with
observations, at the least in the case of M3.

According to such a procedure, Synthetic HB have been
produced, randomly distributing for all the adopted metallicites
800 stars in the He burning phases for selected values of the mean
mass and adopting in all cases a Gaussian dispersion of masses
with a standard deviation $\sigma$=0.02$M_{\odot}$ to account for
the evidence of differential mass loss in the ZAHB progenitor.
As  found in the seminal paper by Rood (1973), such a
value of the standard deviation appears the most adequate to
reproduce the observed distribution of stars along the HB of
typical Galactic globulars (see also Caputo et al. 1984 and Figure
3 in Brocato et al. 2000). However, numerical experiments  show
that the results of the present investigation are not critically
dependent on this assumption, remaining substantially unchanged
when  the adopted dispersion is either increased or decreased by a
factor of two.

For each metallicity and for each mass, we have performed 10
different simulations, thus deriving the mean values and the
corresponding standard deviations for all the relevant quantities.
The complete set of results is available at http:
//www.mporzio. astro. it/ $\sim$marco /sintetici/, where one can find
the data concerning the single HB simulations together with the
plots of the corresponding log$L$, log$T_e$ or $M_V$, $B-V$
diagrams and the period frequency histograms. Table 1 gives
selected results for RR Lyrae rich cases, intended as SHB
simulations where the number of pulsators is $\ge$ 5\% of the global
HB star population.

\begin{table*}
\caption{Selected results of synthetic HB simulations at $Y$=0.23.
Each row gives the metallicity, the mean HB mass,
the HB type, the number of RR Lyrae stars together with their mean
mass, mean visual magnitude and mean fundamentalised period, the
AGB ratio R$^{AGB}$= N(AGB)/N(HB) and the predicted number ratio
of HB to RGB stars more luminous than i) the ZAHB bolometric
magnitude at the RR Lyrae gap (R$^I$), ii) the absolute visual
magnitude of the faintest RR Lyrae pulsator (R$^{II}$) and, iii)
the mean $\langle M_V^{RR}\rangle$ magnitude of the pulsators.
Masses are in solar units. Absolute visual magnitudes of the ZAHB
at log$T_e$=3.83 are given in parentheses below the corresponding
metallicity values.}
\begin{center}
\begin{tabular}{c c r r c c c c c c c }
\hline
\hline\\
$Z$     & $\langle$M$_{HB}\rangle$ & HB type & N(RR)& $\langle$M$_{RR}\rangle$  & $\langle$M$_V^{RR}\rangle$ & $\langle$logP$_f\rangle$&R$^{AGB}$ & R$^{I}$ & R$^{II}$& R$^{III}$  \\
&\\
 \hline

0.0001 &  0.70 & 0.90 &  46.3 &  0.714 &  0.378  & $-$0.233 &  0.11 &1.16 & 1.45&1.56 \\
(0.517 mag) &  0.72 & 0.83 &  80.4 &  0.729 & 0.405  & $-$0.237 &  0.11 &1.14& 1.41&1.50 \\
         &  0.74 & 0.74 & 117.5 &  0.745 &  0.424  & $-$0.245 &  0.11 & 1.14 &1.38& 1.47\\
         &  0.76 & 0.62 & 152.4 &  0.767 &  0.437  & $-$0.259 &  0.11 &1.12 &1.32& 1.41\\
         &  0.78 & 0.44 & 246.9 &  0.791 &  0.451  & $-$0.297 &  0.10 & 1.12&1.32& 1.39\\
         &  0.80 & 0.15 & 401.0 &  0.809 &  0.459  & $-$0.309 &  0.11 & 1.10&1.30& 1.36\\
         &  0.82 &$-$0.09 & 532.3 &  0.822 &  0.460  & $-$0.294 &  0.10 &1.10 &1.29& 1.35\\
         &  0.84 &$-$0.18 & 579.8 &  0.835 &  0.457  & $-$0.266 &  0.10 & 1.10&1.29& 1.35\\
\hline
  0.0003 &  0.66 & 0.91 & 43.3  &  0.678 &  0.435  & $-$0.248 &  0.11 &1.16 &1.40& 1.58\\
(0.607 mag) &  0.68 & 0.78 & 107.2 &  0.696 &  0.490  & $-$0.275 &  0.11 &1.14 &1.35& 1.46\\
         &  0.70 & 0.52 & 228.2 &  0.713 &  0.532  & $-$0.297 &  0.11 &1.14& 1.33& 1.42\\
         &  0.72 & 0.11 & 378.8 &  0.726 &  0.550  & $-$0.293 &  0.12 &1.13 &1.32& 1.38\\
         &  0.74 &$-$0.29 & 394.6 &  0.738 &  0.556  & $-$0.269 &  0.10 & 1.13&1.32& 1.37\\
         &  0.76 &$-$0.60 & 271.8 &  0.747 &  0.555  & $-$0.239 &  0.11 & 1.12&1.30& 1.37\\
         &  0.78 &$-$0.82 & 131.4 &  0.756 &  0.552  & $-$0.212 &  0.11 & 1.11&1.31& 1.35\\
\hline
  0.0006 &  0.64 & 0.89 & 50.8  &  0.664 &   0.478 & $-$0.271 &  0.12 &1.19 &1.41& 1.57\\
(0.630 mag)         &  0.66 & 0.70 & 146.9 &  0.677 &   0.535 & $-$0.288 &  0.12 &1.16 &1.36& 1.47\\
         &  0.68 & 0.28 & 272.9 &  0.688 &   0.568 & $-$0.290 &  0.12 & 1.13&1.32& 1.38\\
         &  0.70 &$-$0.24 & 316.2 &  0.696 &   0.580 & $-$0.278 &  0.12 &1.13 &1.32& 1.37\\
         &  0.72 &$-$0.69 & 188.7 &  0.703 &   0.585 & $-$0.255 &  0.12 & 1.13&1.32& 1.36\\
         &  0.74 &$-$0.91 &  62.3 &  0.710 &   0.588 & $-$0.231 &  0.13 & 1.11&1.32& 1.34\\

\hline

  0.001  &  0.62 & 0.89 &  55.3 &  0.645 &   0.513 & $-$0.270 &  0.11 &1.18& 1.41& 1.57\\
(0.668 mag)         &  0.64 & 0.64 & 155.1 &  0.657 &   0.583 & $-$0.290 &  0.11 &1.16& 1.36& 1.46\\
         &  0.66 & 0.11 & 261.8 &  0.665 &   0.609 & $-$0.293 &  0.12 & 1.13& 1.33& 1.41\\
         &  0.68 &$-$0.44 & 231.7 &  0.672 &   0.623 & $-$0.278 &  0.10 &1.13& 1.33& 1.39\\
         &  0.70 &$-$0.80 & 113.4 &  0.678 &   0.631 & $-$0.261 &  0.11 &1.10& 1.30& 1.35\\
\hline
  0.003  &  0.58 & 0.84 &  57.3 &  0.605 &   0.682 & $-$0.328 &  0.12 &1.14& 1.72& 1.92\\
(0.799 mag)         &  0.60 & 0.40 & 150.0 &  0.611 &   0.733 & $-$0.342 &  0.12 &1.12& 1.69& 1.80\\
         &  0.62 &$-$0.25 & 156.2 &  0.615 &   0.755 & $-$0.337 &  0.13 &1.10& 1.67& 1.77\\
         &  0.64 &$-$0.77 &  81.2 &  0.629 &   0.769 & $-$0.334 &  0.11 &1.10& 1.66& 1.74\\
\hline
  0.006  &  0.56 & 0.79 &  57.0 &  0.584 &   0.787 & $-$0.369 &  0.10 &1.39& 1.84& 2.04\\
(0.906 mag)         &  0.58 & 0.29 & 119.0 &  0.588 &   0.840 & $-$0.381 &  0.12 &1.36& 1.80& 1.91\\
         &  0.60 &$-$0.42 & 111.3 &  0.590 &   0.857 & $-$0.385 &  0.12 &1.34& 1.76& 1.85\\
         &  0.62 &$-$0.86 &  40.6 &  0.592 &   0.877 & $-$0.382 &  0.12 &1.32& 1.74& 1.80\\

\hline \hline
\end{tabular}
\end{center}
\end{table*}

Since this investigation is mainly devoted to the pulsational
behavior of RR Lyrae variables, this issue will be discussed first
in the next sections. However, the adopted SHB procedure offers
the opportunity of clarifying some other relevant
evolutionary parameters, such as the number ratio of RGB to HB stars.
We will devote section 6  to the discussion of such evolutionary
features.

\section{Pulsator predictions}

\begin{figure}
\centering
\includegraphics[width=7.5cm]{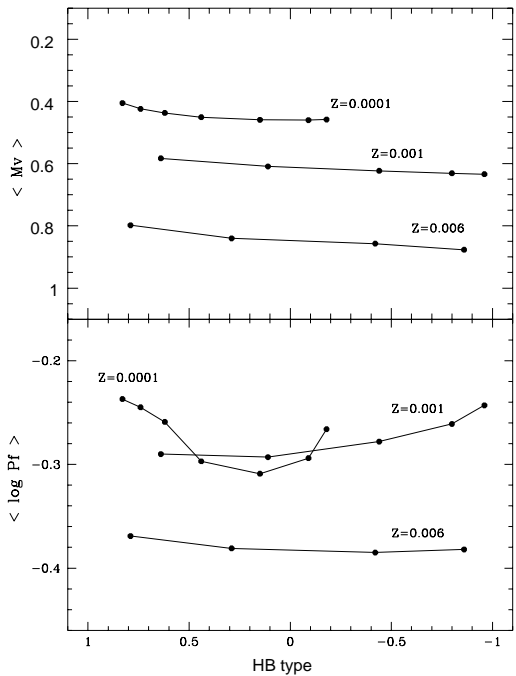}
\caption {Upper panel: Predicted mean visual magnitude of RR
pulsators as a function of the HB type and for the three labelled
metallicities. Lower panel: as in the upper panel, but for the
predicted mean fundamentalised period.} \label{f:fig3}
\end{figure}

For each assumed metal content,  data  in Table 1 show 
some details of the already predicted dependence of the mean RR Lyrae
magnitude on the HB type (see, e.g., Lee, Demarque \& Zinn 1990,
Caputo et al. 1993,
Demarque et al. 2000). As expected, one finds
that this magnitude becomes brighter when the HB morphology moves
from red to blue, i.e., with HB type going from $-$1 to +1. This
is easily understood according to the evidence that in very blue
HBs the RR Lyrae instability strip is populated only with stars in
their later phase of central He burning, crossing the pulsation
region above the ZAHB luminosity level. A similar, but much more
reduced, effect can appear in the case of very red HBs at
$Z$=0.0001: again the strip lacks ZAHB pulsators, being populated
only by stars entering from the red and slightly more luminous
than the local ZAHB level. This is shown in the upper panel in
Fig. \ref{f:fig3}, where the dependence of the mean pulsator
magnitude $\langle M_V^{RR}\rangle$ on the HB type for three
selected metallicities is reported. From data in the figure, one
can also infer the dependence of the same magnitude on the
metallicity.

The lower panel in Fig. 3 shows the predicted dependence of
the mean fundamentalised period on both the HB type and metal
content. To discuss these data, one has to recall that, according
to theory, the period decreases as luminosity decreases and/or
mass or effective temperature increases. As for mean periods, one
finds that at a fixed metallicity when going from blue HB types to
the redder ones, the decrease in luminosity and the simultaneous
increase in mass (see Table 1) both tend to decrease the pulsator
periods. However, data in Fig. 3 show that such a decrease is
successfully counteracted by a temperature effect: when the
population in the cooler portion of the strip begins to dominate,
the mean period starts to increase again, damping the tendency towards a continuous decrease.

Such an effect is more remarkable with $Z\le$ 0.0006 for the
simple reason that at these metallicities the adopted mass
dispersion ($\sigma$ = 0.02$M_{\odot}$) produces the smallest
dispersion of HB temperatures (see Fig. 2) and, thus, the largest
efficiency of the temperature effect. Increasing the metallicity,
the  dispersion in HB temperatures increases, progressively
smoothing away the evidence for the mechanism. Note that  very
metal poor globulars in the Galaxy all have blue HB type, thus
falling along the descending branch of the mean period relation.
Note also that, if Galactic globular cluster ages are comparable
with the Hubble time ($\sim$ 13 Gyr), presently evolving RG are
predicted to be less massive than 0.8 M$_{\odot}$,
allowing only HB types bluer than $\sim$ 0.4 in the most metal-poor clusters. However, the predictions for redder HB
distributions may be of some interest in the case of some metal-poor dwarf spheroidal galaxies in the Local Group which are
characterized by relatively red HBs and are thought to be a sort
of "bridge" between Oosterhoff type I and type II Galactic
globular clusters (see Mateo 1998, Dall'Ora et al. 2003 and
references therein).

\begin{figure}
\centering
\includegraphics[width=8cm]{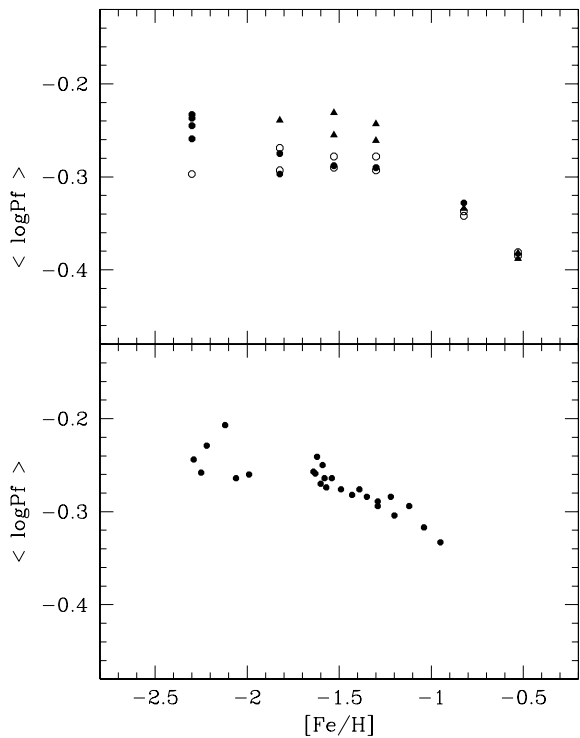}
\caption {Upper panel: Predicted mean fundamentalised periods  as
a function of the hydrogen-to-iron content [Fe/H]=log$Z$+1.7.
Filled circles: HB type $>$ 0.5; open circles: 0.5 $>$ HB type $>$
$-$0.5; triangles: HB type $<$   $-$0.5 . Lower panel: Observed
mean fundamentalised periods for RR Lyrae rich Galactic globular
clusters as a function of the cluster [Fe/H] metallicity (from
Paper I). }

\label{f:fig4}
\end{figure}

Importantly, data in  Fig. 3 confirm the
prediction (Caputo \& Castellani 1975) that
increasing the metallicity  in the low metallicity ($Z\le$ 0.001)
range, the decrease in luminosity (which will induce a decrease in
period) is counteracted by the concurrent decrease in mass. As a
result, over the range  $Z$=10$^{-4}$ to 10$^{-3}$
the mean fundamentalised period is unaffected by
metallicity, at fixed HB morphology. This is not the case for the
higher metal contents, where the effect of luminosity dominates
and mean periods are predicted to decrease when increasing the
metallicity. This is exactly the behavior shown by Galactic
globular clusters as discussed in Paper I. This can be better
appreciated by looking at the upper panel in Fig. \ref{f:fig4},
where we report the predicted mean fundamentalised periods as a
function of the hydrogen-to-iron content [Fe/H]=log$Z$+1.7, in
varying the HB type and  assuming only HB type $\ge$ 0.5 in the
case $Z$=0.0001. Comparison with the observed data, as presented
in Paper I and now displayed in the lower panel of Fig.
\ref{f:fig4}, shows an encouraging similarity, supporting the
capability of the theoretical scenario, as obtained by collating
stellar evolution and pulsation theories, to account for the
actual behavior of RR pulsators in different simple stellar
populations.

We note some relevant
concordances between present predictions and observational data
for Galactic globular clusters as displayed in Table 1 of Paper I.
As an example, by relying on the Harris (1996) metallicity scale, the
two RR rich globulars M5 and M62 have both  metallicity $Z\sim$
10$^{-3}$, quite similar HB type (0.31 and 0.32, respectively)
and, not surprising, quite similar mean fundamentalised periods,
as given by $<$log$P_f>$=$-$0.294 and $-$0.289, respectively. For
this metallicity one can interpolate the data in  Table 1 to
derive for HB type $\sim$ 0.30 a predicted mean fundamentalised
period $<$log$P_f> \sim -$0.290. Going down to $Z\sim$ 10$^{-4}$,
we find M15, with HB type 0.67 and $<$log$P_f>$=$-$0.258, which is
in impressive concordance with the predicted value
$<$log$P_f>-$0.259 at HB type 0.62.

\begin{table}
\caption{The comparison between observed and predicted mean
periods for the sample of RR Lyrae rich Galactic globulars (see
Paper I). Metal abundances and HB type (HB) are from Harris.  }

\begin{center}
\begin{tabular}{l l l r l l}
\hline
\hline\\
NGC /IC &   & [Fe/H] & HB & $\langle$log$P_f\rangle_{ob}$ &
$\langle$log$P_f\rangle_{pr}$ \\
\hline

NGC7078 & M15 & $-$2.25 & 0.67 & $-$0.258 & $-$0.259 \\
NGC4590 & M68 & $-$2.06 & 0.17  &$-$0.264 & $-$0.294 \\
NGC5024 & M53 & $-$1.99 & 0.81 & $-$0.260 & $-$0.259\\
NGC7006 &     &  $-$1.63 & $-$0.28 & $-$0.259 & $-$0.270\\
IC4499 &      &  $-$1.60 & 0.11 & $-$0.270 & $-$0.283\\
NGC6715 & M54 & $-$1.59 & 0.87 &  $-$0.250 & $-$0.259\\
NGC3201 &     &  $-$1.58 & 0.08 & $-$0.264 & $-$0.285\\
NGC5272 & M3 &  $-$1.57 &  0.08 & $-$0.274 & $-$0.285\\
NGC6934 &    &  $-$1.54  & 0.25 & $-$0.264 & $-$0.289\\
NGC6402   &M14 & $-$1.39 & 0.65 & $-$0.276 & $-$0.288\\
NGC5904 & M5  & $-$1.29 & 0.31 & $-$0.294 & $-$0.293\\
NGC6266 & M62 & $-$1.29 & 0.32 & $-$0.289 & $-$0.293\\
\hline \hline
\end{tabular}
\end{center}
\label{t:tab3}
\end{table}

Data in Table \ref{t:tab3} allow the
comparison of present predictions with the observed mean log$P_f$
in the twelve clusters with more than 40 RR Lyrae listed in Table
1 of Paper I. Not surprisingly, predictions for some  clusters
appear less precise, possibly in connection with the uncertainties
in the HB type, with the occurrence of non-negligible statistical
fluctuations   and with the still existing
uncertainties in the evaluation of the actual cluster metallicity
(see, e.g., Kraft \& Ivans 2003, Asplund et al. 2004). From data
in Table 1 one can indeed estimate that an uncertainty of 0.2 dex
in [Fe/H] produces on average an uncertainty of $\sim$ 0.02 in the
mean log$P_f$, and that such an uncertainty on the mean periods
increases to $\sim$0.03 when adding an uncertainty of 0.1 in the
HB type. Bearing this in mind, one finds that all the  RR Lyrae-rich clusters have mean fundamentalised periods which agree with
theoretical predictions within the quoted uncertainty, with the
majority of   clusters, but two (M68 and NGC6934), with
discrepancies lower or of the order of 0.02.

Catelan (2004) has recently discussed the evidence for an
anomalous distribution of RR periods in M3, an occurrence  which
could be at the origin of additional discrepancies,  since our
prediction assumes a smooth distribution of stars within the
instability strip.

\section{The $PL_K$ relations}

The observational evidence has already shown a
tight correlation between fundamental periods and near-infrared
$K$ magnitudes of RR Lyrae stars. Longmore, Fernley \& Jameson
(1986) and Longmore et al. (1990) found that RR Lyrae
stars in Galactic globular clusters follow a well-defined
near-infrared  period-luminosity ($PL_K$) relation, whose slope
appears almost independent of the cluster metallicity. Bono et al
(2001, 2003) have shown that such an occurrence is actually in
agreement with the predictions of current pulsational theory:
making use of suitable analytical approximations of evolutionary
and pulsational results, these authors presented theoretical
predictions on the $PL_K$ in good agreement with observational
constraints. We are now in the position of performing a much more
detailed analysis because the adopted synthetic procedure does not
require any assumption about the predicted distribution of star
masses, luminosities and effective temperatures.

\begin{figure}
\centering
\includegraphics[width=8cm]{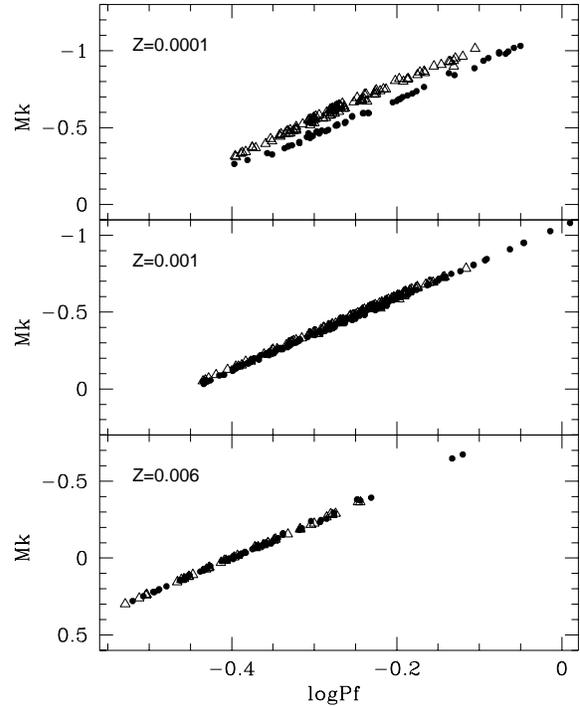}
\caption {Predicted $PL_K$ relation for the three labelled values
of metal content. Each panel shows predictions
concerning the bluest
(circles) and the reddest (triangles) RR Lyrae rich HB type. }
\label{f:fig6}
\end{figure}

Figure 5 shows the synthetic log$P_f$-$M_K$ distribution for three
selected assumptions about the star metallicity presenting, for
each metallicity, predictions concerning the bluest  and the
reddest HB type allowing the occurrence of a substantial fraction
of RR Lyrae pulsators ($\ge$ 5\% of the total number of HB stars).
Data in this figure can be easily understood in terms of the mass
distributions presented in the upper panel of Fig. 2. At the low
metallicity limit $Z$=0.0001, the ZAHB mass is a sensitive
function of the effective temperature and the instability strip
can be populated by a non-negligible range of ZAHB masses. When
evolutionary effects are taken into account, one finds that in the
case of the bluest or the reddest RR Lyrae-rich HB types the
pulsator average mass is around M$\sim$ 0.71 $M_\odot$ and $\sim$
0.84 $M_\odot$, respectively. In Bono et al. (2003) we have
already presented the theoretical prediction

$$M_K \propto -2.102 \log P - 1.753 \log M$$

\noindent from which one derives, for each given log$P$, a
difference between the two cases of $\Delta M_K \sim$ 0.13
mag, as actually
found in the SHB simulations. Increasing the metallicity decreases
the range of masses and the difference is only marginal for
$Z$=0.001 and vanishes for $Z$=0.006.

The exploration of the synthetic results shows that the actual
slope  of the predicted log$P_f$-$M_K$ relation is a bit steeper
than the analytical evaluation at constant mass. We also find that this slope is, within the uncertainties,
not dependent  on the HB type and with a minor dependence on the
metal content. As a whole, putting the $PL_K$ relation in the
usual form

$$M_K  = M_K^{-0.3} +b_K (\log P_f +0.3)$$

\noindent  where $M_K^{-0.3}$ is the $M_K$ value when
log$P_f$=$-$0.3, we derive the predicted values of $M_K^{-0.3}$
and $b_K$ for the various metallicities and HB types, as given in
Table \ref{t:tab2}, where HB types in italics deal with scarce
populations of RR Lyrae pulsators. Note that typical uncertainties
for both $M_K^{-0.3}$ and $b_K$ are $\pm$0.02 mag.

\section{The Wesenheit functions}
\begin{figure}
\centering
\includegraphics[width=8cm]{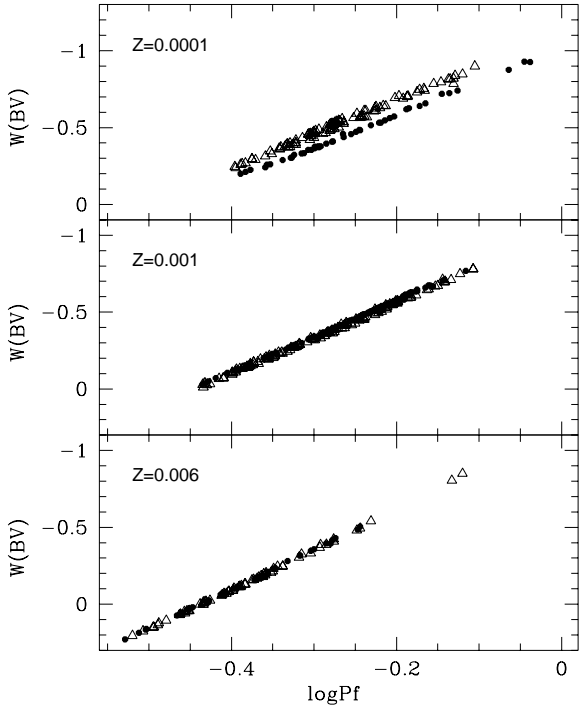}
\caption {Predicted $W(BV)$ functions versus periods for the
labelled values of metal content. Each panel shows predictions
concerning the bluest (triangles) and the reddest (circles) RR
Lyrae rich HB type }
\end{figure}
\begin{figure}
\centering
\includegraphics[width=8cm]{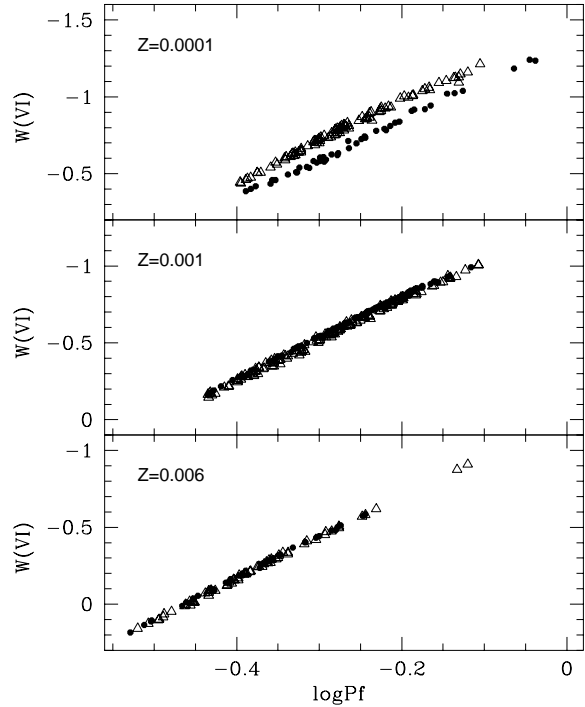}
\caption {As in Fig. 6, but with the $W(VI)$ function.}
\end{figure}

The reddening-free algorithm originally suggested  by van den
Bergh (1975) has proved to be a useful tool for
investigating variable stars (e.g., Classical Cepheids, see Madore
1982), including the RR Lyrae class (Kov\'acs \& Walker 2001,
Paper III). As widely known, the use of Wesenheit functions, such
as $W(BV)=V-3.10(B-V)$, $W(VI)=V-2.54(V-I)$,  has the twofold
advantage of getting rid of reddening but also of
strongly reducing the luminosity dispersion due to the finite
width of the instability strip. In this sense, the
Period-Wesenheit relation ($PW$) works like the $PL_K$, although
on the basis of quite different physical principles. This matter
has been discussed in Paper III within a purely pulsational
scenario and it seems interesting to improve that discussion on
the basis of the present synthetic procedure.

\begin{table*}
\caption{Slope and zero point (in magnitudes) of predicted $PL_K$
and $PW$ relations for various metallicity and HB type  with
significant numbers of RR Lyrae pulsators ($\ge$ 5\% the total
number of HB stars), except the HB types in italics. }

\begin{center}
\begin{tabular}{l r r r r  c c c}
\hline
\hline\\
Z&HBtype& $b_K$ &  $M_K^{-0.3}$ & $b_{BV}^W$ & W$_{BV}^{-0.3}$ &
$b_{VI}^W $ & W$_{VI}^{-0.3}$  \\

\hline
\\[1pt]

0.0001&{\it 0.99} & $-$2.30 &  $-$0.439 & $-$2.15 & $-$0.359&      $-$2.54&  $-$0.532\\
&0.90 & $-$2.30 &  $-$0.455 & $-$2.15 & $-$ 0.370&      $-$2.54&  $-$0.594\\
&0.83   & $-$2.30 &  $-$0.465 & $-$2.15 & $-$0.378 &     $-$2.54 & $-$0.608\\
&0.62&    $-$2.30 &  $-$0.498&  $-$2.15 & $-$0.409 &     $-$2.54&  $-$0.649\\
&0.44 &   $-$2.30 &  $-$0.516 & $-$2.15&  $-$0.427 &     $-$2.54&  $-$0.669\\
&0.15 &   $-$2.30  & $-$0.532 & $-$2.15&  $-$0.442 &     $-$2.54 & $-$0.689\\
&$-$0.09  & $-$2.30  & $-$0.543&  $-$2.15 & $-$0.451  &    $-$2.54 & $-$0.704\\
&$-$0.18 &  $-$2.30 &  $-$0.553 & $-$2.15 & $-$0.459 &     $-$2.54 &$-$0.716\\

\hline
0.0003& {\it 0.97} & $-$2.34 &  $-$0.399 & $-$2.25 & $-$0.332&      $-$2.61&  $-$0.530\\
&0.91  &  $-$2.34  & $-$0.404&  $-$2.25 & $-$0.337 &     $-$2.61 &$-$0.545\\
&0.78 &   $-$2.34 &  $-$0.416 & $-$2.25 & $-$0.348 &     $-$2.61 & $-$0.566\\
&0.52  &  $-$2.34 &  $-$0.427  &$-$2.25&  $-$0.358 &     $-$2.61 & $-$0.583\\
&0.11  &  $-$2.34 &  $-$0.435 & $-$2.25 & $-$0.365 &     $-$2.61&  $-$0.596\\
&$-$0.29  & $-$2.34 &  $-$0.442  &$-$2.25 & $-$0.371 &     $-$2.61 & $-$0.606\\
&$-$0.60 &  $-$2.34 &  $-$0.449 & $-$2.25&  $-$0.378 &     $-$2.61&  $-$0.614\\
&$-$0.82 &  $-$2.34&   $-$0.458&  $-$2.25&  $-$0.386&      $-$2.61&  $-$0.524\\

\hline
0.0006& {\it 0.97} & $-$2.34 &  $-$0.382 & $-$2.30 & $-$0.333&      $-$2.63&  $-$0.513\\
&0.89  &  $-$2.34 &  $-$0.384 & $-$2.30&  $-$0.338 &     $-$2.63& $-$0.530\\
&0.70  &  $-$2.34&   $-$0.389  &$-$2.30 & $-$0.343 &     $-$2.63&  $-$0.543\\
&0.28  &  $-$2.34 &  $-$0.394 & $-$2.30  &$-$0.348 &     $-$2.63 & $-$0.555\\
&$-$0.24 &  $-$2.34 &  $-$0.397 & $-$2.30&  $-$0.351 &     $-$2.63&  $-$0.561\\
&$-$0.69 &  $-$2.34 &  $-$0.404 & $-$2.30 & $-$0.359 &     $-$2.63 & $-$0.569\\
&$-$0.91&   $-$2.34 &  $-$0.406&  $-$2.30  &$-$0.361&      $-$2.63 & $-$0.572\\

\hline
0.001& {\it 0.97} & $-$2.34 &  $-$0.357 & $-$2.33 & $-$0.327&      $-$2.65&  $-$0.502\\
&0.89  &  $-$2.34  & $-$0.359 & $-$2.33 & $-$0.331  &    $-$2.65 &$-$0.508\\
&0.64 &   $-$2.34&   $-$0.357 & $-$2.33 & $-$0.329 &     $-$2.65 & $-$0.517\\
&0.11& $-$2.34 &  $-$0.362&  $-$2.33& $-$0.334&     $-$2.65 & $-$0.525\\
&$-$0.44 &$-$2.34 &$-$0.362&  $-$2.33 & $-$0.334  &    $-$2.65 & $-$0.525\\
&$-$0.80 &$-$2.34 &$-$0.367 &$-$2.33  &$-$0.340 &     $-$2.65 & $-$0.536\\
&$-$0.96& $-$2.34 &$-$0.370 &$-$2.33& $-$0.346 &     $-$2.65 & $-$0.539\\

\hline

 0.003& 0.9 to $-$0.9 &   $-$2.34&   $-$0.280 & $-$2.45 & $-$0.327  &    $-$2.66 & $-$0.459\\
\hline
 0.006 & 0.9 to $-$0.9 &  $-$2.41   &$-$0.242 & $-$2.60 & $-$0.356 &     $-$2.68&$-$0.438\\

\hline \hline
\end{tabular}
\end{center}
\label{t:tab2}
\end{table*}

The predicted $W(BV)$ and $W(VI)$ functions are plotted in
Fig. 6 and Fig. 7, respectively, as a function of period
and for the labelled choices on the pulsator metallicity.
As in the $PL_K$ case, one finds that the dispersion of pulsator
masses can play a relevant role only at the lower metallicities,
where the predicted distributions are significantly
dependent on the HB type. On the contrary, for metal content larger or of the order of $Z$=0.001 one expects a tight and univocal $PW$
relation. Again one finds that the slope of the
predicted relation is independent of the HB type, being only
marginally dependent on the pulsator metallicity.

In all cases,
putting the theoretical predictions in the form

$$W = W^{-0.3} + b^W (\log P_f +0.3)$$

\noindent we derive the values listed in Table \ref{t:tab2} for
both the $W(BV)$ and $W(VI)$ functions. Uncertainties on these
values are of the same order of magnitude ($\pm$0.02 mag) as in
the case of the $PL_K$ relation.

As a conclusion, we predict that,
at least in moderately metal-rich
clusters, both the $PL_K$ and the $PW$ relations  are able to
provide strong constraints to the cluster distance moduli
independent of the HB type. Concerning the reliability of the
presented theoretical result,
we notice that the slope of both the
$PL_K$ and $PW$ relations are dependent only on the relation
between periods and stellar structural parameters, such as $L$ and
$Te$, which represent a well-established and firm result of the
current pulsational scenarios.
Conversely, the zero points of the predicted
relations depend on the assumption about the luminosity
of HB models and on the adopted bolometric corrections,
thus requiring firmer theoretical or empirical
constraints. Here, we recall that all through
this paper we made use
of HB luminosities following
the theoretical dependence on
metallicity, as predicted by stellar evolution theory,
and calibrated to fit the observed period distribution
of RR Lyrae stars in M3.

\section{Evolutionary parameters}

Since the pioneering paper by Iben (1968), the number ratio R
between HB and RG stars more luminous than the HB luminosity level
has been widely used as an evolutionary indicator of the star
original He content. The theoretical calibration of this parameter
can be performed by simply evaluating the ratio of theoretical
lifetimes in the two quoted evolutionary phases. However,  the
lifetimes in the central He burning phase is a function of the
mass of HB stellar structures, depending on the amount of mass
lost by their RG progenitors. Since HB stars in actual globular
clusters cover a non-negligible range of masses, the calibration
of R  requires an appropriate mean of the HB evolutionary
times, especially when hot long-living HB stars are present (see
lower panel in Fig. 2).

Since the first exhaustive analysis by Buzzoni et al. (1983), the
calibration of R has been the subject of several investigations
attempting to take into consideration the dependence of He burning
lifetimes on the effective temperature of HB structures. However,
as already stated in Zoccali et al. (2000), a correct measure of
the absolute He content on the basis of the R-parameter would
require the construction, for any given cluster, of synthetic CMD
which properly reproduce the distribution of stars along the HB.
The synthetic procedure adopted in this paper indeed allows  a
straightforward evaluation of this parameter  by randomly
populating both the RG and HB phases for any given assumption on
the cluster metallicity and for any given assumption about the HB
masses and, thus, about the cluster HB type.

\begin{figure}
\centering
\includegraphics[width=8cm]{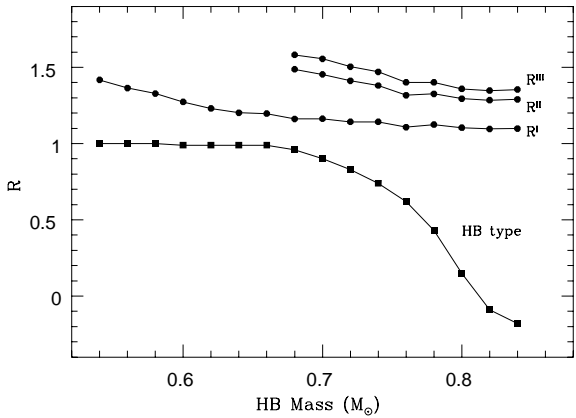}
\caption {Predicted values of the number ratios R$^I$, R$^{II}$
and R$^{III}$ (see text) versus the mean HB mass, at $Z$=0.0001.
For comparison, the HB type is also reported
(squares).} \label{f:fig5}
\end{figure}

According to the already-described procedure, we have evaluated
the predicted values of R  all along our sample of synthetic HB
simulations. However, it is not intended to produce firm predictions
on this parameter. As a fact, it has already brought to light
(see, e.g., Brocato et al. 1998, Cassisi et al. 1998) the evidence
that HB lifetimes, and thus the theoretical calibration of R, are
sensitively affected by -at least- current uncertainties on the
nuclear cross section for the He burning reaction
$^{12}$C($\alpha, \gamma$)$^{16}$O. Thus, the given evaluations of
R are mainly intended to investigate the sensitivity of this
parameter to the HB type at the various metallicities, putting on a
more robust basis the results already presented on the matter by
several authors (Caputo, Martinez Roger \& Paez 1987, but see also
Zoccali et al. 2000, Cassisi, Salaris \& Irwin 2003).

However, one has also to account for the several slightly different
definitions of R adopted in the literature as far the HB
luminosity level is concerned. As an example, Zoccali et al (2000)
assumed as a reference the visual magnitude of the lower envelope of
the observed HB distribution at the temperature of the RR Lyrae
gap, taken as representative of the ZAHB luminosity. In the same
year, Sandquist (2000)  took the bolometric
luminosity corresponding to the average $V$ magnitude of HB stars
at the same temperatures. We note  the already
advanced warning about the use of the lower envelope of RR
luminosity as indicative of the ZAHB luminosity (Ferraro et al.
1999, but see also Paper II). To investigate the consequences of
different assumptions about the reference luminosity levels, Table
1 gives results for three different approaches, where R$^I$
represent the number ratio of HB to RG stars more luminous than
the predicted ZAHB luminosity at logT$_e$ = 3.85,
R$^{II}$ the same ratio but for  RG stars
more luminous than the faintest
RR luminosity level and R$^{III}$ for RG stars
more luminous than the RR Lyrae mean
magnitude.

As expected,  the three values are in increasing order, for
the very simple reason that the reference luminosity level is
progressively increasing from R$^I$ to R$^{III}$, driving a
consequent decrease in the number of RG stars. Data in Table 1
should be read bearing in mind that, according to current
calibrations, a variation in R by $\Delta$R$ \sim$ 0.1, if
interpreted in terms of original He, would imply  a $\Delta$Y a
bit larger than 0.01 (Buzzoni et al 1983, but see also Caputo,
Martinez Roger \& Paez 1987, Bono et al. 1995). One finds  that,
in the extreme case, for each given metallicity the HB type can
move the R$^{II}$ value up to $\Delta$R $\sim$ 0.12, i.e., with a
variation larger than predicted for R$^I$, where only the effects
of HB evolutionary lifetimes are at work. In passing, note that
the sudden increase of R values at the large metallicity follow
the decreasing luminosity of the RGB bump (see the discussion in
Zoccali et al. 2000).
\begin{figure}
\centering
\includegraphics[width=8cm]{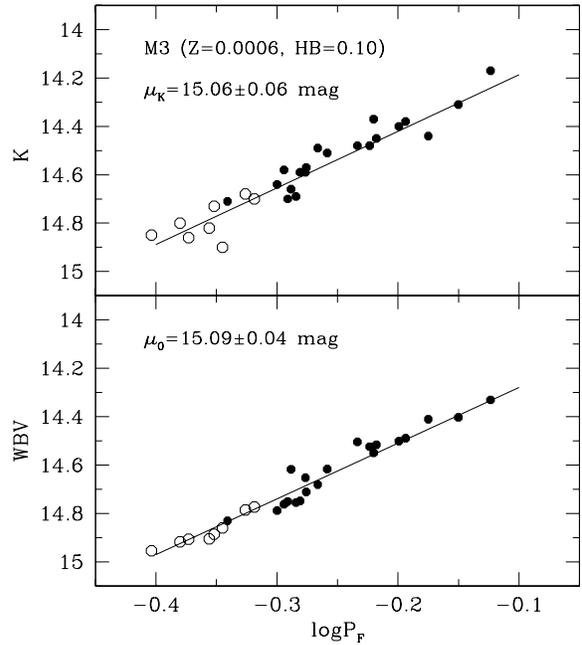}
\caption {Predicted $PL_K$ and $PW(BV)$ relations in comparison
with observed data in the globular cluster M3.} \label{f:fig9}
\end{figure}

\begin{figure}
\centering
\includegraphics[width=8cm]{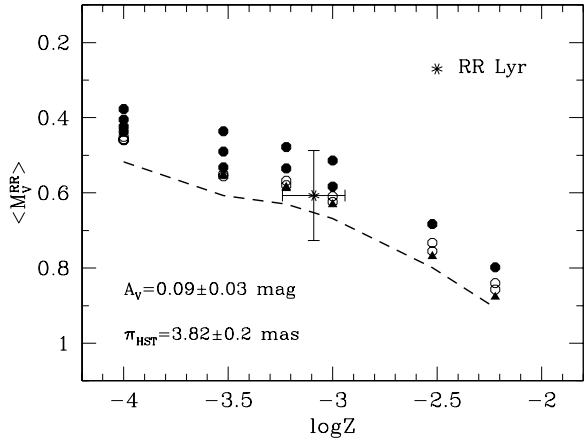}
\caption {Predicted mean magnitude of RR Lyrae stars versus
metallicity for different HB types, with symbols as in Fig.4. The
dashed line depicts the ZAHB behavior. The location of the
prototype RR Lyr is shown according to the labelled HST
trigonometric parallax and visual extinction.} \label{f:fig10}
\end{figure}

To allow a more general discussion, Fig. \ref{f:fig5} shows
theoretical predictions with $Z$=0.0001, as produced  from the
entire sample of synthetic HBs., i.e., not only for RR
Lyrae rich cases. The adopted metallicity follows from the
evidence that with such a choice one is maximizing the increase of
HB time when increasing HB effective temperatures. One notices
that in the range of RR Lyrae rich HB types, the effects of
lifetimes are of minor relevance. However, going toward hotter HB,
the effects of HB lifetime grows, as expected, producing a
variation of R$^I$ up to $\sim$ 0.3. Data in Table 1 shows that the difference between
R$^{II}$ and R$^{III}$ is increasing with increasing 
metallicity. This is because synthetic HB computations
predict that the width in
luminosity is increasing with metallicity, thus increasing the
difference between the lower luminosity level  and
$\langle$M$_V^{RR}\rangle$. In this sense, the theory appears in
agreement with the empirical correlation presented by Sandage (1987).

\section{Discussion and final remarks}

The main purpose of this paper is to show that synthetic  HB
procedures can be usefully adopted to explore not only
globular cluster
CM diagrams,  but also to predict several relevant features of
cluster RR Lyrae variables. We have presented the architecture and
the results of the adopted procedure, showing in particular the
encouraging concordance of several predictions with the observed
distribution of fundamentalised periods at the various
metallicities. In the following papers, we plan to make use of such a
theoretical tool to discuss the variable populations in single
well observed Galactic and extragalactic clusters.

However,  we give a brief  comparison of our
predictions with available data for the well studied globular
cluster M3 for which, following the recent metallicity scale by
Kraft \& Evans (2003), we adopt $Z$=0.0006. All the predicted relations are based on
"static" magnitudes, whereas observed data deal with quantities
averaged over the pulsation cycle. This issue has been discussed
in Paper II and Paper III and we wish only to recall that in the
case of $K$ magnitudes the discrepancy between static and mean
values is at the most of the order of 0.01 mag, whereas it
increases moving to visual and to blue photometric bands, as well
as from symmetric (low amplitude) to asymmetric (high amplitude)
light curves. On this basis, the observed mean $K$ magnitudes can
be directly compared with the predicted $PL_K$ relation, whereas
the observed $W(BV)$ and $W(VI)$ quantities need to be first
corrected for the amplitude effect (see Paper II and Paper III).

Figure 9 shows the comparison of M3 observed data with present
theoretical predictions (solid lines) both for the $PL_K$ (upper
panel) and the $PW$ relation (lower panel).  We regard the close
agreement between empirical and predicted slopes as satisfactory
evidence, whereas we notice that a discrepancy of only 0.03 mag in
the derived cluster distance modulus  is not only well within our
estimated uncertainty but, possibly, also within the empirical
uncertainty of the mean $K$ magnitudes, as obtained in 1990 by
Longmore et al. New and better $K$ data would thus be needed to
explore in more detail the internal consistency of the presented
scenario.

We show in Fig. 10 the predicted mean
magnitudes $\langle M_V^{RR}\rangle$ listed in Table 1, as derived
for the various RR Lyrae rich HB types, versus the metal content.
For comparison, the level of ZAHB at the RR Lyrae gap
is also drawn with a dashed line. Looking at the results plotted
in Fig. 10, one should agree that it is quite risky to predict a
{\it unique} $\langle M_V^{RR}\rangle$-log$Z$ relation independent
of the HB type, a result which appears quite consistent with the
significantly different empirical relations that have appeared in the
relevant literature. Moreover, one should also consider that our
predictions hold for a constant helium content $Y$=0.23 and that
accounting for $\Delta Y/\Delta Z\sim$ 2.5 will produce not
negligible effects on the results at $Z$=0.003 and 0.006, leading
to magnitudes brighter by $\sim$ 0.04 and 0.08 mag, respectively.

However,  our theoretical
predictions agree with the absolute magnitude of the prototype RR
Lyr, as estimated from the recent HST trigonometric parallax and
visual extinction estimate (see Benedict et al. 2002). We are facing a  relevant agreement between theory and firm observational constraints, disclosing the progresses performed  since the first unsuccessful but pioneering attempts by Rood (1973).

The computed synthetic HB contains of course much more information
than that summarized in this paper. Detailed
results of the single synthetic  models have been  made
available at the already quoted www site,  together with the
corresponding plots of the log$L$, log$Te$ and $V, B-V$ predicted
diagrams for the whole HB and the upper portion of the RG branch
and with  Tables summarizing the whole data set.

\begin{acknowledgements}
We thank our anonymous referee for helpful comments and
suggestions. This research has made use of NASA's Astrophysics
Data System Abstract Service and SIMBAD database operated at CDS,
Strasbourg, France. This work was partially supported by MURST
(PRIN2002, PRIN2003, PRIN2004)
\end{acknowledgements}

\end{document}